# Investigation of heat treatment condition, ethylene glycol and deionized water effects on the phase structure of $Ba_{0.6}Sr_{0.4}TiO_3$ nanoparticles produced by the sol-gel method

Mohammadali Ajam[1], Abolghasem Nourmohammadi[2] *, S. Mohammad Hasan Feiz[1]

[1]Department of Physics, University of Isfahan, Isfahan, Iran

[2]Department of Nanotechnology Engineering Faculty of Advanced Sciences and Technologies, University of Isfahan, Isfahan, Iran

**Abstract**

$Ba_{0.6}Sr_{0.4}TiO_3$ nanoparticles were synthesized by the sol-gel method. The effects of the heat treatment conditions, ethylene glycol (EG) and deionized water (DW) on their phase structure and formation of the undesirable barium carbonate ($BaCO_3$) phase were investigated. It was found that increasing the temperature of the samples caused an increase in the peak intensities of perovskite structures, and a decrease in the intensities of the $BaCO_3$ peaks. Moreover, omitting EG, the sol stabilizer, and water, the solvent, decreased the peak intensities of BaCO3. Decreasing the heating rate caused better burnout of the contained organic materials and, as a result reduced the peak intensity of BaCO3. But, it was impossible to remove the BaCO3 completely up to 850 ºC. However, single-phase perovskite barium strontium titanate nanoparticles were obtained at 750 ºC by using an acetic acid based precursor sol through decreasing the gel heating rate, and omitting EG and DW.

* Corresponding author. Email: a.nourmohammadi@sci.ui.ac.ir

## 1. Introduction

Lead-based perovskite ferroelectric materials such as $Pb(Zr_xTi_{1-x})O_3$ or PZT are widely used in industry because of their superior ferroelectric and piezoelectric properties. However, they are harmful and will be banned in the near future. The RoHS European directive (Restriction on Hazardous substances) has prohibited lead in electric and electronic applications since July 2006. However, no lead-free ferroelectric ceramic has better or equal piezoelectric performance compared with PZT[1]. For this reason, a lot of research works are being performed in order to modify the ferroelectric and piezoelectric properties of lead-free provskite materials. Barium strontium titanate or BST ($Ba_xSr_{1-x}TiO_3$) is a good candidate for lead-free piezoelectrics. This material possesses a high dielectric constant and a composition-dependent Curie temperature, and hence, shows potential applications in piezoelectric and pyroelectric sensors, dynamic random access memories (DRAM), tunable filters and microwave phase shifters[2, 3, 4].

Several groups of researcher have used different chemical composition to modify electrical and electronical properties of $Ba_xSr_{1-x}TiO_3$ material according to Ba content in the solid solution ($0 \leq x \leq 1$)[5] in order to obtain high crystallinity, controlled particle size and shape, etc. Experience shows that the best room temperature dielectric properties of BST are observed for the compositions around x=0.7[6, 7, 8]. To achieve the desired properties and practical applications, the quality of the BST powders are very important, which depends strongly on their synthesis. It is known that fine, homogeneous, and dispersive nanosized powders are necessary for the development of a uniform microstructure with desired properties.

There are different methods to prepare high crystalline BST nanopowders including gel combustion [3], polymeric precursor (pechini process)[9], solid-state reaction[10], hydrothermal reaction[11] and sol-gel method[12]. Among them, the sol-gel process has significant advantages over other methods because of the purity and homogeneity of the final material, precise control of the stoichiometry and ease of doping[13].

As reported by some researchers[9], sol-gel preparation of BST nanopowders has several restrictions; 1) Application of alkoxide titanium precursors requires a closed container filled with inert gas atmosphere. 2) The presence of the organic components causes to form barium carbonate materials during decomposition of the organic materials. Barium carbonate can be formed easily, but it is highly stable. Decomposition of this by product ($BaCO_3$) usually requires heat treatment at elevated temperatures[14].

For removing barium carbonate and facilitate the powder synthesis, the researchers here have investigated the following experience for obtaming (BST) pure phase:

1. The effect of deionized water solely.
2. The effect of simultaneous using ethylene glycol and deionized water.
3. Influence of the absence of both deionized water and ethylene glycol.
4. The effects of heat treatment conditions.

To the best our knowledge, the combination of these effects have not been considered by other researches on the crystallization of BST nanopowders with the pure perovskite phase. In present work, we investigated $Ba_{0.6}Sr_{0.4}TiO_3$ material at different processing conditions as we will describe in the experimental procedure section. We selected x=0.6, which is quite close the peaking BST composition of x=0.7, in order to avoid any perturbation due to cubic-tetragonal phase transformation. The highly crystalline BST nanopowders with the pure perovskite phase were successfully prepared at 750 ℃.

## 2. Experimental

### 2.1 Sol-gel processing

Titanium (IV) isopropoxide (Ti(OCH(CH$_3$)$_2$)$_4$Merck, 98% purity), barium acetate (Ba(CHCOO)$_2$, Merck, 99.99% purity), strontium acetate (Sr(CHCOO)$_2$, Merck, 99.9% purity) were chosen as source materials. Pure glacial acetic acid was used as the modifier, ethylene glycol and acetyl acetone as the sol stabilizer, deionized water and 2-methoxyethanol were selected as the solvents.

Then, strontium acetate and barium acetate were added in a hot mixture of acetic acid and deionized water. Then, titanium (IV) isopropoxide was mixed with 2-methoxyethanol and the prepared mixture was added into the above mentioned solution along with continuous stirring. Then, the ethylene glycol and acetyl acetone were added into the mixture for the stability of the sol. The value of $p$H was adjusted, using glacial acetic acid, to remain at 4. The solutions were mixed and stirred at a temperature of about 65 °C for 0.5 h to obtain a clear solution. The concentration of the final solution was adjusted at 0.5M.

### 2.2 Heat treatment and characterization methods

Three types of precursor sols were prepared in this study which are labeled as A, B, C (table 1). In order to prepare the gel samples, the precursor sols were heated at 80 °C for several hours. Then, the prepared gels were dried at 200 °C for 48 hours. In table 1, samples $A_1$-$A_7$ contain both EG and DW. However, they have prepared in different heat treatment conditions. Both samples $B_1$ and $B_2$ have DW but without any EG. They have heat treated at 750 °C and 850 °C, respectively. Finally, sample C has neither EG nor DW. The latter sample is heat treated at the lowest temperature to reach the pure perovskite BST (see section 3.2).

Shape, size and crystal structure of the prepared nanoparticles were investigated using a transmission electron microscope (TEM) system model Philips CM30 with electron diffraction facilities. Then, the prepared nanopowders were analyzed by X-Ray diffraction (XRD) method employing a Bruker-d8 advance model using Co Kα radiation (1.789 Å) in order to identify the crystal structure of the calcinated BST nanoparticles and calculation of their average crystallite size.

## 3. Results and discussion

### 3.1 Heat treatment process

The produced BST nanopowders were heat treated under different conditions to reach the nanocrystals of the desired barium strontium titanate material.

The results of heat treatment of $Ba_{0.6}Sr_{0.4}TiO_3$ nanopowders have come in table 1. As it is seen in this table, the samples heat treated with the lowest heating rate have bright white color, which indicate existence of less organic components due to more decomposition of organics (samples $A_1$-$A_3$ in table 1). Hence, a 2.5 ºC/min heating rate was utilized in the rest of our study.

### 3.2 XRD patterns

Fig1 shows the XRD patterns of the powder samples prepared using different precursor sols, after heat treatment at different conditions (see table 1). The initial XRD patterns show formation of the perovskite barium strontium titanate nanopowders, along with barium carbonate, after calcination. According to Fig 1, the intensity of the peaks due to the undesirable $BaCO_3$ phase reduces with increasing the heat treatment temperature (Figs 1a – 1c). It should be noted that it was impossible to get rid of the peaks due to this undesirable phase in the samples containing EG and DW, even at 850 ºC. Their intensity decreases drastically in the samples without any EG, although small peaks due to $BaCO_3$ are observed after heat treatment at 850 ºC. It was found in this study that, in the absence of both DW and EG, pure perovskite phase BST nanopowders were formed at a lower temperature,750 ºC, (sample C in Table 1 and Fig 1). All the diffraction peaks due to sample C correspond to the perovskite BST phase, according to the JCPDS card No 34-0411. The average size of the nanocrystals was calculated using the Willamson-Hall plot (not shown here), which came out to be about 31nm.

### 3.3 TEM results

TEM microstructural and electron diffraction analyses were performed in order to investigate size, morphology and phase composition of the prepared nanoparticles. To confirm our XRD results, our optimized sample (BST nanopowder prepared in the absence of EG and DW in the precursor sol) was investigated.

As shown in Fig 2, after heat treatment at 750 °C for 2 hours, the prepared nanoparticles mostly have size of around 33nm. The average nanoparticle size which is measured by using TEM images is often larger than the average size, determined by XRD for the nanocrystals. The reason is that some nanoparticles are polycrystal and compose several nanocrystals. Fig 3 indicates that most of the nanoparticles produced in our research work are single crystal because their size is around 33nm, which is quite close to the average nanocrystallite size, calculated by our XRD data (31 nm).

Fig 3 shows the electron diffraction pattern of the nanoparticles in Fig 2. As shown in this figure, single phase perovskite crystallization of BST is obtained, because all the diffraction points correspond to BST, according to the aforementioned standard JCPDS card. It is demonstrated that low temperature single phase crystallization of BST nanoparticles has occurred at 750°C whereas crystallization of BST nanomaterials has already been reported above 850°C [5, 15]. Moreover, Fig 3 indicates that our method has been a suitable one for preparing high quality crystalline BST nanoparticles.

## 4. Conclusions

In this research work, the effects of heat treatment, ethylene glycol and deionized water on the phase structure of $Ba_{0.6}Sr_{0.4}TiO_3$ nanoparticles were investigated. Our experimental results showed that omitting ethylene glycol and deionized water in the BST sol preparation leads to a $Ba_{0.6}Sr_{0.4}TiO_3$ pure phase at a lower temperature. It was observed that the heating rate causes to burn better the organic components in the samples. So, it results in bright white color in our samples. In addition, the effects of heat treatment and heating rate on our samples, in the range of 650 ºC-850 ºC were considered. Thus, it was observed that selecting a 2.5 ºC/min heating rate and a 750 ºC calcination temperature in absence of ethylene glycol and deionized water results in formation of pure BST perovskite phase.

Sol composition and sol preparation process have key contribution in the phase composition of the final BST powder. In the presence of EG and DW, it was impossible to obtain pure perovskite BST nanoparticles blow 850 ºC. However, in the absence of EG and DW formation of single phase BST nanopowders was obtained at 750 ºC.

The average size of the nanocrystals, fabricated at 750 ºC, was calculated by the Willamson-Hall method came out to be about 31nm. Our transmission electron microscopy (TEM) investigations confirmed formation of single crystal BST nanoparticles with around 33nm size. The average nanoparticle size which is measured by using TEM images is often larger than the average size, measured by XRD for the nanocrystals. Hence, most of the nanoparticles produced in our research work are possibly single crystal because their size is quite close to the average nanocrystallite size, calculated by our XRD data (31nm).

**Tables**

**Table 1.** Different $Ba_{0.6}Sr_{0.4}TiO_3$ nanopowders samples and heat treatment conditions investigated in this research work.

| Hample | Heating rate (ºC/min) | Temperature (ºC) | Time (h) | Color Sample | EG | DW |
|---|---|---|---|---|---|---|
| $A_1$ | 5 | 650 | 2 | White | Yes | Yes |
| $A_2$ | 10 | 650 | 2 | Gray | Yes | Yes |
| $A_3$ | 20 | 650 | 2 | Darker gray of $A_2$ | Yes | Yes |
| $A_4$ | 5 | 700 | 2 | white | Yes | Yes |
| $A_5$ | 2.5 | 700 | 2 | Bright White | Yes | Yes |
| $A_6$ | 2.5 | 750 | 2 | Bright White | Yes | Yes |
| $A_7$ | 2.5 | 850 | 2 | Bright White | Yes | Yes |
| $B_1$ | 2.5 | 750 | 2 | Bright White | No | Yes |
| $B_2$ | 2.5 | 850 | 2 | Bright White | No | Yes |
| C | 2.5 | 750 | 2 | Bright White | No | No |

**Figure captions**

**Fig 1.** X-ray diffraction patterns of the different prepared samples in table1.

**Fig 2.** The TEM image of our optimized $Ba_{0.6}Sr_{0.4}TiO_3$ nanoparticles (sample C, table 1).

**Fig 3.** The electron diffraction pattern of our optimized $Ba_{0.6}Sr_{0.4}TiO_3$ nanoparticles in Fig 2.

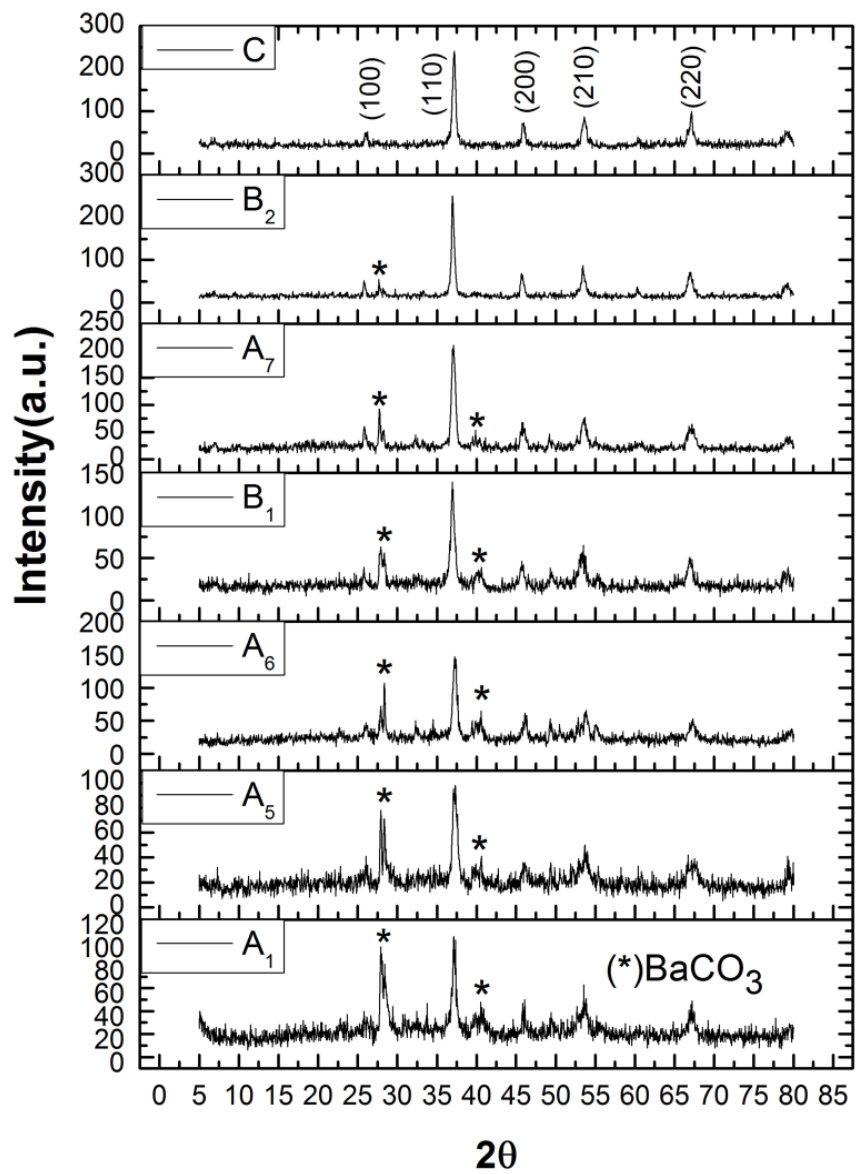

Figure 1

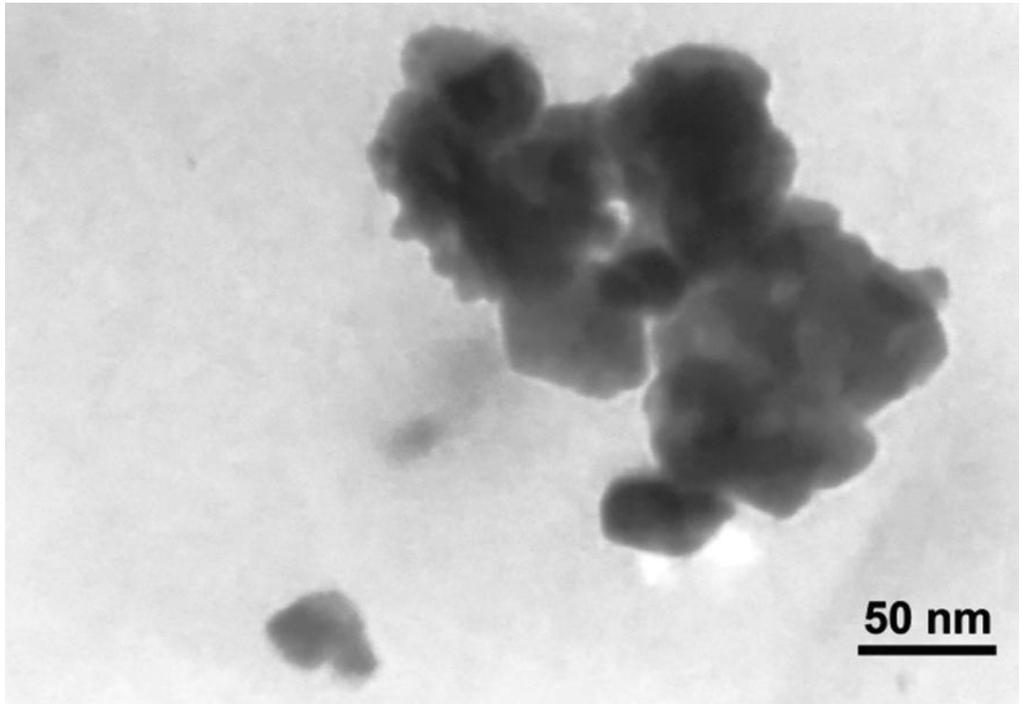

Figure 2

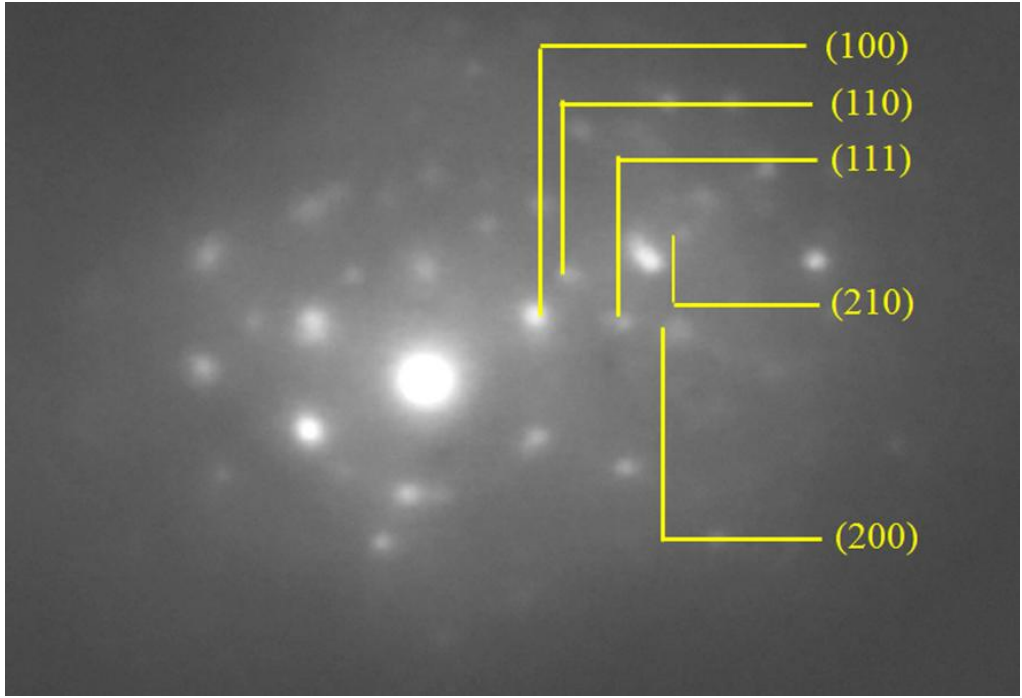

Figure 3